\documentclass[english,aps,prper,reprint,superscriptaddress,floatfix,longbibliography]{revtex4-1}   
\usepackage{lipsum}
\usepackage[T1]{fontenc}	
\usepackage[latin9]{inputenc}	
\usepackage{geometry}		
\usepackage{graphicx}
\usepackage{color}
\usepackage[above,below]{placeins}	
\usepackage{mathptmx}
\usepackage[colorlinks=true,allcolors=blue]{hyperref}
\usepackage{multirow}
\usepackage{rotate}

\usepackage{lineno,xcolor}
\usepackage{amsmath,amssymb}
\usepackage{array}
\usepackage[caption = false]{subfig}

\begin{document}
\title{Developing a reasoning inventory for measuring physics quantitative literacy} 

\author{Trevor I.\ Smith}
\affiliation{Department of Physics \& Astronomy, Rowan University, 201 Mullica Hill Rd., Glassboro, NJ 08028, USA}\affiliation{Department of STEAM Education, Rowan University, 201 Mullica Hill Rd., Glassboro, NJ 08028, USA}
\author{Suzanne W.\ Brahmia}
\affiliation{Department of Physics, University of Washington, Box 351560, Seattle, WA 98195-1560, USA}

\author{Alexis Olsho}
\affiliation{Department of Physics, University of Washington, Box 351560, Seattle, WA 98195-1560, USA}

\author{Andrew Boudreaux}
\affiliation{Department of Physics and Astronomy, Western Washington University, 516 High St, Bellingham, WA 98225, USA}

\author{Philip Eaton}
\affiliation{Department of Physics, Montana State University, Bozeman, Montana 59717, USA}

\author{Paul J.\ Kelly}
\affiliation{Department of Physics \& Astronomy, Rowan University, 201 Mullica Hill Rd., Glassboro, NJ 08028, USA}

\author{Kyle J.\ Louis}
\affiliation{Department of Physics \& Astronomy, Rowan University, 201 Mullica Hill Rd., Glassboro, NJ 08028, USA}\affiliation{Department of STEAM Education, Rowan University, 201 Mullica Hill Rd., Glassboro, NJ 08028, USA}

\author{Mitchell A.\ Nussenbaum}
\affiliation{Department of Physics \& Astronomy, Rowan University, 201 Mullica Hill Rd., Glassboro, NJ 08028, USA}

\author{Louis J.\ Remy}
\affiliation{Department of Physics \& Astronomy, Rowan University, 201 Mullica Hill Rd., Glassboro, NJ 08028, USA}

\begin{abstract}
In an effort to improve the quality of citizen engagement in workplace, politics, and other domains in which quantitative reasoning plays an important role, Quantitative Literacy (QL) has become the focus of considerable research and development efforts in mathematics education. QL is characterized by sophisticated reasoning with elementary mathematics. In this project, we extend the notions of QL to include the physics domain and call it Physics Quantitative Literacy (PQL). We report on early stage development from a collaboration that focuses on reasoning inventory design and data analysis methodology for measuring the development of PQL across the introductory physics sequence. We have piloted a prototype assessment designed to measure students' PQL in introductory physics: Physics Inventory of Quantitative Literacy (PIQL). This prototype PIQL focuses on two components of PQL:  proportional reasoning, and reasoning with negative quantities. We present preliminary results from approximately 1,000 undergraduate and 20 graduate students.
\end{abstract}

\maketitle

\section{Introduction}
\label{sec:intro}
The development of mathematical reasoning skills is an important goal in many introductory physics courses, particularly those geared toward students majoring in physics and other physical science and engineering fields. Previous research has shown that students' development of Physics Quantitative Literacy (PQL)---the ability to reason mathematically in the context of physics---is often less than desired \cite{Brahmia2017c}; however, few studies have rigorously examined the development of PQL over time or how this might vary across student populations. We have begun developing the Physics Inventory of Quantitative Literacy (PIQL) to address the need for a valid and reliable assessment instrument for measuring students' PQL across the undergraduate physics curriculum. 

Enhancing PQL has the potential to strengthen students' knowledge of mathematics \cite{Thompson2010,Ellis2007}, better prepare them for future demands to think quantitatively \cite{Caballero2015}, and promote increased equity and inclusion in physics instruction \cite{Brahmia2017a,Boaler2015}. At its core, quantitative literacy (QL) involves blended use of mathematical concepts and procedures. Both everyday sense-making and workplace performance rely on QL, and physics is ideally positioned to help students develop these skills. We have developed an 18-question prototype PIQL that focuses on two specific elements of PQL: reasoning with ratios and proportions, and reasoning about negative quantities. 


The use of ratios and proportions to describe systems and characterize phenomena is a hallmark of expertise in STEM fields, perhaps especially in physics. Boudreaux, Kanim, and Brahmia identify a set of reasoning subskills to provide a more fine-grained analysis of proportional reasoning, and they isolate college students' specific proportional reasoning difficulties based on assessment items designed to span the proportional reasoning space \cite{Boudreaux2015}. The items are categorized into six subskills, which overlap with the early work of Arons, as `underpinnings' to success in introductory physics \cite{Arons1983}.

Unlike physics experts, novices often have difficulty understanding the many roles signed numbers can play in physics contexts. Brahmia and Boudreaux constructed physics assessment items based on the natures of negativity from mathematics education research \cite{Vlassis2004} and administered them to introductory physics students \cite{Brahmia2017c}. They find that students have trouble reasoning about signed quantity in several contexts typically found in the undergraduate curriculum (e.g., negative work, or negative direction of electric field) \cite{Brahmia2017a,Brahmia2016b}. Bajracharya, Wemyss, and Thompson report that students struggle to make meaning of negative area under a curve in physics contexts \cite{Bajracharya2012}. Hayes \& Wittmann report students having difficulty with the use of negative signs to attribute direction to acceleration in a functional representation \cite{Hayes2010}. All of these studies reveal that signed quantities, and their various meanings in introductory physics, present cognitive difficulties for students that many don't reconcile before completing the introductory sequence. 

Because QL is ubiquitous throughout the undergraduate physics curriculum, we expect that there may be interaction effects between students' mathematical reasoning skills and their abilities to apply them in multiple physics (and non-physics) contexts. As such, the protoPIQL includes multiple physical contexts for each component of PQL.

We administered the 18-question protoPIQL to students in three different introductory physics courses at a large public research university in the northwestern United States. Students completed the protoPIQL during the first week of class in each of the three quarter-long introductory physics courses: mechanics ($N=459$), electricity \& magnetism ($N=328$), and waves, optics, and thermodynamics ($N=317$). For comparison purposes, we also administered the negativity questions to 22 graduate students in physics; we consider the graduate students to be a nearly ``expert'' population, as they have continued beyond their undergraduate studies in physics. We present the results from this initial implementation, looking both at the instrument as a whole, and specific items that yield particularly interesting results.

\section{Whole-Test Results}
\label{sec:results}
The protoPIQL consists of 18 multiple-choice questions: 8 permit only a single response and 10 allow students to select multiple responses. For our first round of analyses, students were scored as being correct on a question if they selected all correct responses and no incorrect responses. For all data analysis we include only students who answered at least 2/3 of the questions (1,076 out of 1,104 respondants). 

\begin{figure}[tb]
	\includegraphics[width = 0.4\textwidth]{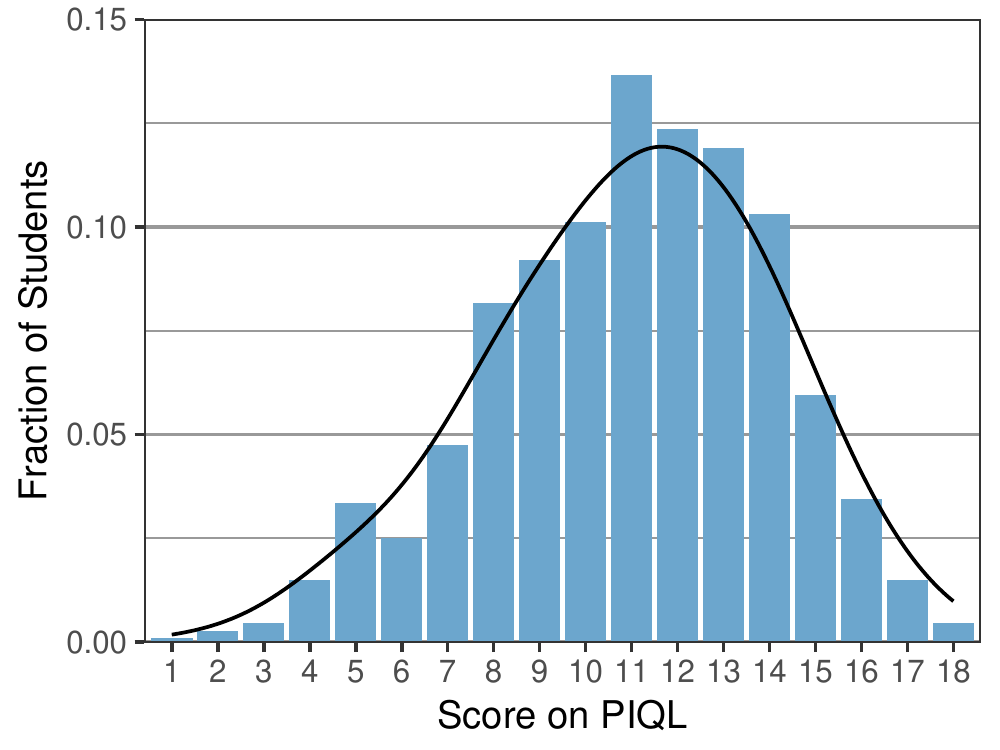}
    \caption{Distribution of scores on the protoPIQL ($N\nobreak=\nobreak1,076$).}
    \label{piqlScores}
\end{figure}

Overall, scores are fairly normally distributed (see Fig.\ \ref{piqlScores}), with an average (mean, median, and mode) of 11 out of 18 correct, a standard deviation of 3.0, and small but negative values of both skewness and kurtosis (-0.3 and -0.2, respectively). The internal reliability is Cronbach's $\alpha = 0.67$, which is below the commonly accepted thresholds of 0.8 for making measurements of individuals and 0.7 for making measurements of groups \cite{Doran1980}; however, this may be due to the protoPIQL explicitly measuring two different constructs: proportional reasoning, and negativity \cite{Adams2010}.

We use classical test theory (CTT) to evaluate the quality of each question in terms of its difficulty and its discrimination. In CTT, the difficulty is defined as the fraction of students who answer a particular question correctly. Difficulty ranges from 0 to 1 with lower values indicating questions that are harder for students. CTT discrimination is the difference between the difficulty of each question for high-scoring and low-scoring students (upper vs.\ lower 27\% with regard to test score) \cite{wiersma1985educational}. Discrimination ranges from 0 to 1, with higher values indicating a question that is answered differently for high-scoring and low-scoring students.


\begin{figure}[tb]
	\centering
    \includegraphics[width = 0.4\textwidth]{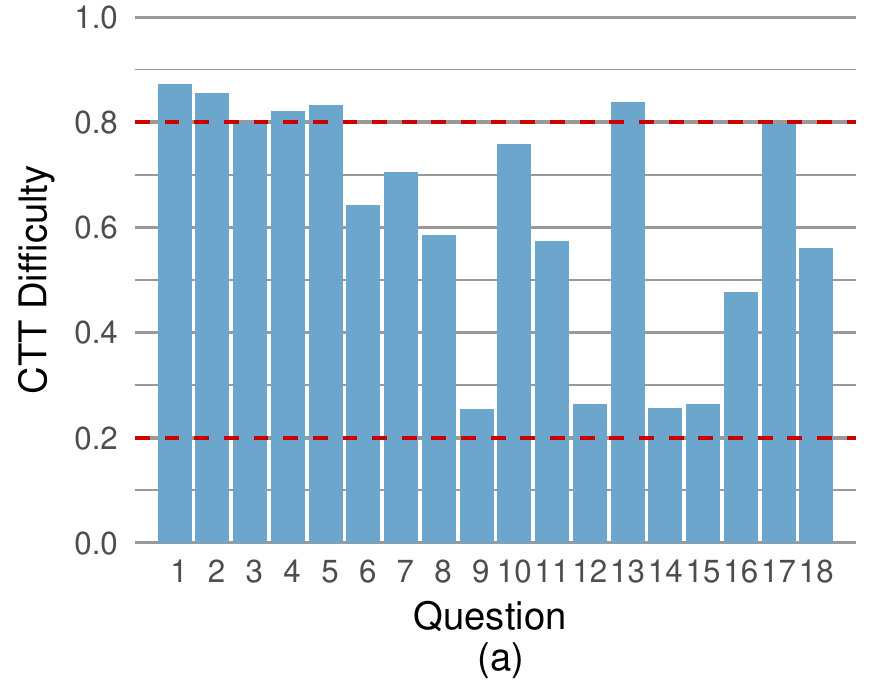}


	\includegraphics[width = 0.4\textwidth]{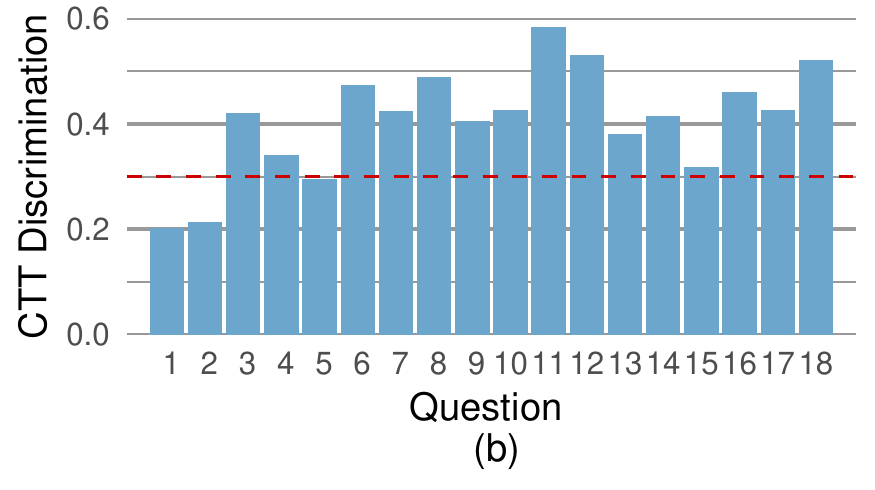}

    
    \caption{Classical Test Theory results: (a) Difficulty, and (b) Discrimination for each question. Red lines show lower (and upper) thresholds for desired values.}
\label{ctt}
\end{figure}

Figure \ref{ctt} shows the CTT difficulty and discrimination for each question. The average CTT-difficulty is 0.62, and 13 of the 18 questions fall within the generally accepted range of $0.2\leq D \leq0.8$, with the remaining 5 being too easy \cite{Doran1980}. Of note in Fig.\ \ref{ctt}(a) is that most questions have difficulties above 0.5, but there are four questions with difficulty values near 0.25. The average CTT item discrimination between the upper and lower 27\% of the class is 0.41, and 15 of the questions were above the common threshold of 0.3 with one more being just below the threshold (0.296 for Q5) \cite{Doran1980}. Of note in Fig.\ \ref{ctt}(b) is that no question achieves a discrimination above 0.6, which is often considered to be highly discriminating. It should also be noted that the maximum possible item discrimination in CTT is dependent on the item difficulty for $D<0.27$ or $D>0.73$ \cite{Doran1980}; for example, the maximum discrimination for Q1 ($D=0.872$) is 0.426, so the minimum threshold for discrimination should be adjusted to $0.3\times0.426=0.142$. With this modification, all questions have acceptable discrimination values (above 0.3 times the maximum value based on the difficulty).

Some questions have difficulty values above the common threshold of 0.8, but we do not necessarily consider these questions to be useless. We want some questions to be fairly easy; some of the quantitative literacy skills we are trying to assess are typically taught in early high school, and the fact that 15--20\% of students in university-level physics courses still have difficulty with these questions is notable. Of course, we don't want to have too many questions with high difficulty parameters or else the assessment loses its power to accurately measure students' reasoning abilities. We use these results as part of our overall evaluation to determine which questions should be kept, modified, or discarded. Other considerations will include a balance in content so that one component of PQL is not featured more heavily than others, and external expert opinion that an item is appropriate and important for our target population.




    

As mentioned previously, the protoPIQL contains single-response questions (SR) in which, students may only choose one response) and multi-response questions (MR), in which students may choose any combination of responses. The MR questions can be further subdivided into those with only a single correct response (MRS) and those with multiple correct responses (MRM).\footnote{Our decision to use the multiple-response, single-correct (MRS) format is based on the desire to see whether or not students will choose incorrect responses in combination with a single correct response. This also allows us to measure students' likelihood to choose various combinations of incorrect responses.} There are three MRM questions: 9, 12, and 15. Comparing this list to the results in Fig.\ \ref{ctt}(a) shows a clear trend: all of the MRM questions are among the most difficult questions. Question 14 is the only MRS question with a difficulty level similar to the MRM questions; Q14 asks students about the meaning of a negative component of an electric field $E_x = -10$ N/C --- a task that may be beyond the abilities of students who have not yet taken a course in electricity and magnetism (70\% of the data set). 

Table \ref{cttGrouped} shows the average difficulty and discrimination for these three groups of questions. These results indicate that the MRS questions are statistically similar to the SR questions in terms of CTT difficulty (both with averages well above the ideal mean of 0.5 \cite{Doran1980}), and that the MRM questions are much harder. This may be a result of students being much less likely to correctly guess the answer to a MRM question given that they must choose the correct combination of responses. Another factor here is that the correct responses to MRM questions tend to involve different aspects of physical quantities. For example, on Q12 students must recognize that negative work indicates that the direction of a component of a force is opposite the direction of the displacement (response C) and that negative work indicates that the energy of the system is decreasing (response E). This may be more complex than recognizing that a negative acceleration means that a component of acceleration is in the negative direction (as on Q11, a MRS question), and previous research indicates that few students use both vector reasoning and scalar reasoning when answering these types of questions \cite{Brahmia2017c}. Given these large differences in CTT difficulty values, we may be reaching (or exceeding) the limit on CTT's usefulness, or we may have a situation in which CTT analyses are not appropriate. Interestingly, there are no statistically significant differences between the CTT discrimination values for SR, MRS, and MRM questions. 

\begin{table}[bt]
\caption{CTT results, separated by question type: single response (SR), multiple response with single correct answer (MRS), and multiple response with multiple correct answers (MRM). Values indicate the mean, and uncertainty is the standard error.}
	\begin{ruledtabular}
		\begin{tabular}{llll}
		&SR&MRS&MRM\\
        \hline
        Difficulty&$0.74\pm 0.04$&$0.64\pm 0.08$&$0.261\pm 0.003$\\
        Discrimination&$0.39\pm 0.04$&$0.42\pm 0.04$&$0.42\pm 0.06$\\
		\end{tabular}
	\end{ruledtabular}
    \label{cttGrouped}
\end{table}

\begin{figure}[tb]
	\includegraphics[width = 0.45\textwidth]{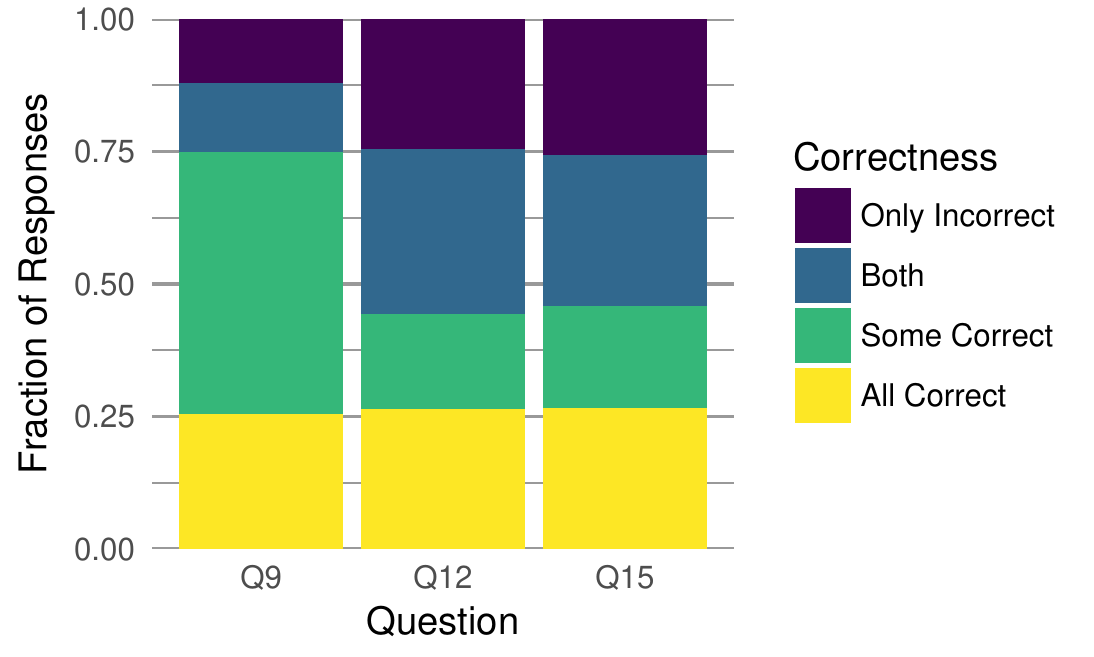}
    \caption{MRM question results with a 4-tiered correctness scale.}
    \label{mrm}
\end{figure}

The big issue here seems to be that students need to correctly choose multiple answers, and each may correspond with a different piece of knowledge. To examine these questions in more depth we have categorized student responses using a multilevel correctness scale: selecting all correct answers (All Correct), selecting at least one correct answer (Some Correct), selecting correct and incorrect answers (Both), or selecting exclusively incorrect answers (Only Incorrect). Figure \ref{mrm} shows the results on the MRM questions using these mutually exclusive levels. The fraction of All Correct responses to each question is consistent with Fig.\ \ref{ctt}(a) at about 25\% for each question, but the distribution of partially correct answers shows that at least 75\% of students are choosing one of the correct responses, even if they also choose an incorrect response. Q9 has the largest fraction of students (about 50\%) choosing some (but not all) correct responses, with Q12 and Q15 each having about 20\% of students in this category. This is notable because Q9 is the only question with three correct responses and two incorrect responses (one of which is a none-of-the-above option), and Q12 and Q15 each have two correct responses and three incorrect responses (with no none-of-the-above option). These results provide evidence that analysis methods beyond the traditional correct/incorrect dichotomy should be explored to fully represent students' understanding of these topics.

\section{Identifying troublesome questions: Charge transfer between comb and hair}
\label{sec:q15}
As mentioned above, the MRM questions yield some interesting results in terms of distributions of partially correct answers. We have found that data from Q15 is particularly interesting with regard to who chooses which responses. The text of Q15 is shown in Fig.\ \ref{Q15}. Answering Q15 correctly requires students to know that the charges that are able to move from one object to another are negative electrons (response A), and to recognize that the net charge has both a magnitude and a sign (similar to a net force having a magnitude and a direction) and the ``size'' of the net charge depends only on the magnitude of the charge: going from zero to nonzero indicates an increase in the magnitude (response E).

Questions 14--16 all involve E\&M topics: negative component of an electric field (Q14), negative charge (Q15), and negative potential difference (Q16). We expected that students who had completed an E\&M course would do better on these questions. Our data indicate that this is true for Q14 and Q16, with statistically significant differences revealed by chi-square analyses ($p < 0.001$), and students in the beginning of their waves/optics/thermo course being more likely to be correct than either students at the beginning of mechanics or students at the beginning of E\&M. But for Q15 our data indicate that students in the beginning of E\&M are more likely to be completely correct than students at the beginning of mechanics, but students at the beginning of waves/optics/thermo are not any better: completing an E\&M course does not affect students' likelihood of being correct on Q15. 

\begin{figure}[tb]
\framebox{\parbox{0.45\textwidth}{\raggedright

Valeria combs her hair, and as a result the net charge on the comb goes from 0 to -5 C.
Consider the following statements about this situation. Select the statement(s) that \textbf{must be true}. \textbf{\textit{Choose all that apply.}}

	\begin{enumerate}
	\setlength{\itemsep}{0.0ex}
	\renewcommand{\labelenumi}{\alph{enumi}.}
	\item \textbf{Negative charge was added to the comb.}
	\item Charge was taken away from the comb.
	\item All of the electric charge in the comb is negative.
	\item The net charge on the comb is smaller after Valeria combs her hair.
	\item \textbf{The net charge on the comb is larger after Valeria combs her hair.}
	\end{enumerate}
    }}
\caption{Question 15: correct responses are bold.}
\label{Q15}
\end{figure}

\begin{figure}[tb]
	\includegraphics[width = 0.45\textwidth]{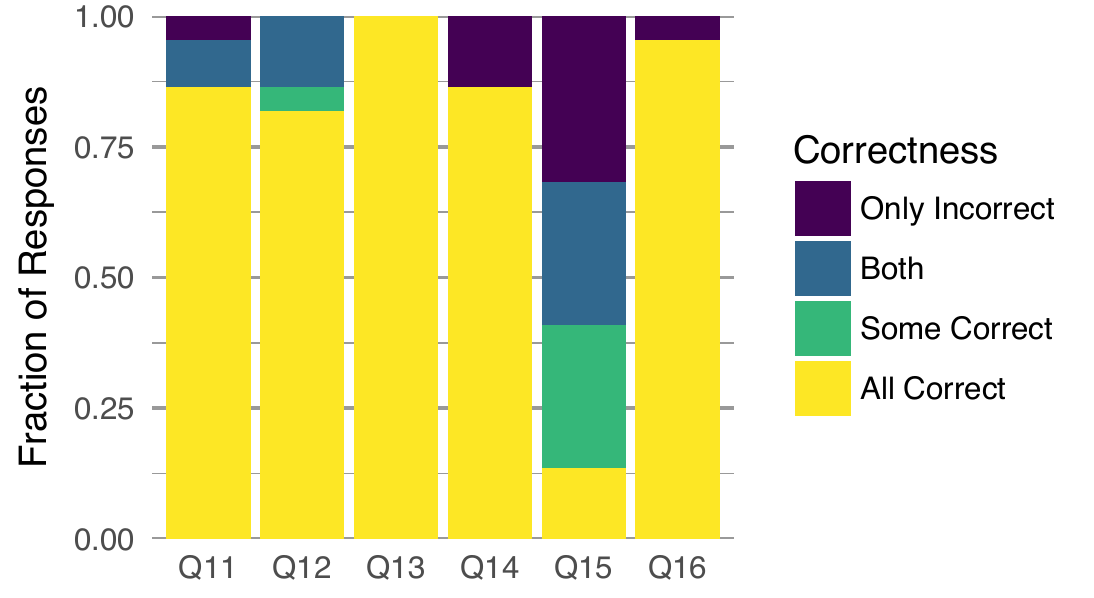}
    \caption{Graduate student results using the 4-tiered correctness scale ($N=22$). Note: Questions 11, 13, 14, and 16 each only have a single correct answer, so the ``Some Correct'' category does not apply for those.}
    \label{grad}
\end{figure}

Moreover, we gave six of the protoPIQL questions to 22 graduate students in physics: results are shown in Fig.\ \ref{grad}. For five of the questions (11--14, and 16) at least 18 students were completely correct, but only three of the 22 students correctly answered Q15. Six students were partially correct (choosing either A or E), and another six students chose a combination of correct and incorrect responses (the most popular being A and D). Q15 gives these students trouble in a way that is different from other questions. These results, coupled with those from the introductory student population, suggest that there is something different (and potentially problematic) about Q15.

Informal feedback from research participants and our colleagues suggests that some of the issues with Q15 may come from different interpretations of the term ``net charge.''  The charge on the comb goes from 0 C to -5 C, which means that the comb starts with a balance of positive and negative charges and ends with a surplus of one type of charge, so the net charge is larger (response E). In this case, the surplus of charge is negative, which leads some to claim that the net charge is smaller because negative numbers are smaller than zero. We disagree with this interpretation because the labels of the types of charge are arbitrary: they could be called North and South as is typical for magnetic poles, but for historical reasons and mathematical ease we use the labels positive and negative. Some colleagues have suggested using the term ``magnitude of the net charge,'' but the ``magnitude'' is an unsigned quantity. A net charge has both magnitude and sign (or type), so specifically asking about the magnitude would limit our abilities to measure how students interpret the negative sign. In an effort to get more input and try alternate wording (such as ``amount of unbalanced charge''), we have begun interviewing introductory students to determine the ways in which they interpret the various questions in the protoPIQL. This process is vital for the validation of the PIQL for measuring physics students' PQL \cite{Adams2010}.

\section{Summary and Future Directions}
The results from the protoPIQL are a promising start for developing a valid and reliable assessment for measuring undergraduate physics students' PQL. In addition to classical test theory, future analyses will include item response theory (both with dichotomously scored data and with the nominal response model), exploratory and confirmatory factor analysis (to ensure that the assessment reflects the desired balance in components of PQL), and additional interviews with both undergraduate students and physics faculty to ensure that the PIQL measures what we want it to measure and that the items are appropriate and important for our target population. Our goal is to achieve good psychometric test parameters while including both breadth and depth of content, and to create an assessment that students can complete in about 30 minutes. We will also explore novel scoring techniques to value growth in students' reasoning skills, not just mastery of the topic. Results from individual questions that appear somewhat anomalous (like those presented from Q15) will help identify the frontiers of future research into physics students' quantitative literacy and reveal topics that are persistently difficult for advanced undergraduate (and graduate) students.

\begin{acknowledgments}
We are grateful to the instructors who administered the \linebreak[2] protoPIQL questions in their classes, and to the students who participated in the research. This work was supported by NSF grants DUE-1832880, DUE-1832836, and DUE-1833050.
\end{acknowledgments}

\bibliography{piqlbib}

\begin{thebibliography}{16}%
\makeatletter
\providecommand \@ifxundefined [1]{%
 \@ifx{#1\undefined}
}%
\providecommand \@ifnum [1]{%
 \ifnum #1\expandafter \@firstoftwo
 \else \expandafter \@secondoftwo
 \fi
}%
\providecommand \@ifx [1]{%
 \ifx #1\expandafter \@firstoftwo
 \else \expandafter \@secondoftwo
 \fi
}%
\providecommand \natexlab [1]{#1}%
\providecommand \enquote  [1]{``#1''}%
\providecommand \bibnamefont  [1]{#1}%
\providecommand \bibfnamefont [1]{#1}%
\providecommand \citenamefont [1]{#1}%
\providecommand \href@noop [0]{\@secondoftwo}%
\providecommand \href [0]{\begingroup \@sanitize@url \@href}%
\providecommand \@href[1]{\@@startlink{#1}\@@href}%
\providecommand \@@href[1]{\endgroup#1\@@endlink}%
\providecommand \@sanitize@url [0]{\catcode `\\12\catcode `\$12\catcode
  `\&12\catcode `\#12\catcode `\^12\catcode `\_12\catcode `\%12\relax}%
\providecommand \@@startlink[1]{}%
\providecommand \@@endlink[0]{}%
\providecommand \url  [0]{\begingroup\@sanitize@url \@url }%
\providecommand \@url [1]{\endgroup\@href {#1}{\urlprefix }}%
\providecommand \urlprefix  [0]{URL }%
\providecommand \Eprint [0]{\href }%
\providecommand \doibase [0]{http://dx.doi.org/}%
\providecommand \selectlanguage [0]{\@gobble}%
\providecommand \bibinfo  [0]{\@secondoftwo}%
\providecommand \bibfield  [0]{\@secondoftwo}%
\providecommand \translation [1]{[#1]}%
\providecommand \BibitemOpen [0]{}%
\providecommand \bibitemStop [0]{}%
\providecommand \bibitemNoStop [0]{.\EOS\space}%
\providecommand \EOS [0]{\spacefactor3000\relax}%
\providecommand \BibitemShut  [1]{\csname bibitem#1\endcsname}%
\let\auto@bib@innerbib\@empty
\bibitem [{\citenamefont {Brahmia}(2017)}]{Brahmia2017c}%
  \BibitemOpen
  \bibfield  {author} {\bibinfo {author} {\bibfnamefont {Suzanne~White}\
  \bibnamefont {Brahmia}},\ }\bibfield  {title} {\enquote {\bibinfo {title}
  {Negative quantities in mechanics: a fine-grained math and physics conceptual
  blend?}}\ }in\ \href@noop {} {\emph {\bibinfo {booktitle} {Physics Education
  Research Conference 2017}}},\ \bibinfo {series and number} {PER Conference}\
  (\bibinfo {address} {Cincinnati, OH},\ \bibinfo {year} {2017})\ pp.\ \bibinfo
  {pages} {64--67}\BibitemShut {NoStop}%
\bibitem [{\citenamefont {Thompson}(2010)}]{Thompson2010}%
  \BibitemOpen
  \bibfield  {author} {\bibinfo {author} {\bibfnamefont {Patrick~W}\
  \bibnamefont {Thompson}},\ }\bibfield  {title} {\enquote {\bibinfo {title}
  {Quantitative reasoning and mathematical modeling},}\ }\href@noop {}
  {\bibfield  {journal} {\bibinfo  {journal} {New perspectives and directions
  for collaborative research in mathematics education}\ ,\ \bibinfo {pages}
  {33}} (\bibinfo {year} {2010})}\BibitemShut {NoStop}%
\bibitem [{\citenamefont {Ellis}(2007)}]{Ellis2007}%
  \BibitemOpen
  \bibfield  {author} {\bibinfo {author} {\bibfnamefont {Amy~B.}\ \bibnamefont
  {Ellis}},\ }\bibfield  {title} {\enquote {\bibinfo {title} {The influence of
  reasoning with emergent quantities on students' generalizations},}\ }\href
  {\doibase 10.1080/07370000701632397} {\bibfield  {journal} {\bibinfo
  {journal} {Cognition and Instruction}\ }\textbf {\bibinfo {volume} {25}},\
  \bibinfo {pages} {439--478} (\bibinfo {year} {2007})}\BibitemShut {NoStop}%
\bibitem [{\citenamefont {Caballero}\ \emph {et~al.}(2015)\citenamefont
  {Caballero}, \citenamefont {Wilcox}, \citenamefont {Doughty},\ and\
  \citenamefont {Pollock}}]{Caballero2015}%
  \BibitemOpen
  \bibfield  {author} {\bibinfo {author} {\bibfnamefont {Marcos~D}\
  \bibnamefont {Caballero}}, \bibinfo {author} {\bibfnamefont {Bethany~R}\
  \bibnamefont {Wilcox}}, \bibinfo {author} {\bibfnamefont {Leanne}\
  \bibnamefont {Doughty}}, \ and\ \bibinfo {author} {\bibfnamefont {Steven~J}\
  \bibnamefont {Pollock}},\ }\bibfield  {title} {\enquote {\bibinfo {title}
  {Unpacking students' use of mathematics in upper-division physics: where do
  we go from here?}}\ }\href {http://stacks.iop.org/0143-0807/36/i=6/a=065004}
  {\bibfield  {journal} {\bibinfo  {journal} {European Journal of Physics}\
  }\textbf {\bibinfo {volume} {36}},\ \bibinfo {pages} {065004} (\bibinfo
  {year} {2015})}\BibitemShut {NoStop}%
\bibitem [{\citenamefont {Brahmia}\ and\ \citenamefont
  {Boudreaux}(2017)}]{Brahmia2017a}%
  \BibitemOpen
  \bibfield  {author} {\bibinfo {author} {\bibfnamefont {Suzanne}\ \bibnamefont
  {Brahmia}}\ and\ \bibinfo {author} {\bibfnamefont {Andrew}\ \bibnamefont
  {Boudreaux}},\ }\bibfield  {title} {\enquote {\bibinfo {title} {Signed
  quantities: Mathematics based majors struggle to make meaning},}\ }in\
  \href@noop {} {\emph {\bibinfo {booktitle} {Proceedings of the 20th Annual
  Conference on Research in Undergraduate Mathematics Education}}},\ \bibinfo
  {series and number} {The Special Interest Group of the Mathematical
  Association of Americ},\ \bibinfo {editor} {edited by\ \bibinfo {editor}
  {\bibfnamefont {Aaron}\ \bibnamefont {Weinberg}}, \bibinfo {editor}
  {\bibfnamefont {Chris}\ \bibnamefont {Rasmussen}}, \bibinfo {editor}
  {\bibfnamefont {Jeffrey}\ \bibnamefont {Rabin}}, \bibinfo {editor}
  {\bibfnamefont {Megan}\ \bibnamefont {Wawro}}, \ and\ \bibinfo {editor}
  {\bibfnamefont {Stacy}\ \bibnamefont {Brown}}}\ (\bibinfo {address} {San
  Diego, CA},\ \bibinfo {year} {2017})\BibitemShut {NoStop}%
\bibitem [{\citenamefont {Boaler}(2015)}]{Boaler2015}%
  \BibitemOpen
  \bibfield  {author} {\bibinfo {author} {\bibfnamefont {Jo}~\bibnamefont
  {Boaler}},\ }\href@noop {} {\emph {\bibinfo {title} {Mathematical mindsets:
  Unleashing students' potential through creative math, inspiring messages and
  innovative teaching}}}\ (\bibinfo  {publisher} {John Wiley \& Sons},\
  \bibinfo {year} {2015})\BibitemShut {NoStop}%
\bibitem [{\citenamefont {Boudreaux}\ \emph {et~al.}(2015)\citenamefont
  {Boudreaux}, \citenamefont {Kanim},\ and\ \citenamefont
  {Brahmia}}]{Boudreaux2015}%
  \BibitemOpen
  \bibfield  {author} {\bibinfo {author} {\bibfnamefont {Andrew}\ \bibnamefont
  {Boudreaux}}, \bibinfo {author} {\bibfnamefont {Stephen}\ \bibnamefont
  {Kanim}}, \ and\ \bibinfo {author} {\bibfnamefont {Suzanne}\ \bibnamefont
  {Brahmia}},\ }\bibfield  {title} {\enquote {\bibinfo {title} {Student
  facility with ratio and proportion: Mapping the reasoning space in
  introductory physics},}\ }\href@noop {} {\bibfield  {journal} {\bibinfo
  {journal} {arXiv preprint arXiv:1511.08960}\ } (\bibinfo {year}
  {2015})}\BibitemShut {NoStop}%
\bibitem [{\citenamefont {Arons}(1983)}]{Arons1983}%
  \BibitemOpen
  \bibfield  {author} {\bibinfo {author} {\bibfnamefont {Arnold~B.}\
  \bibnamefont {Arons}},\ }\bibfield  {title} {\enquote {\bibinfo {title}
  {Student patterns of thinking and reasoning},}\ }\href {\doibase
  10.1119/1.2341417} {\bibfield  {journal} {\bibinfo  {journal} {The Physics
  Teacher}\ }\textbf {\bibinfo {volume} {21}},\ \bibinfo {pages} {576--581}
  (\bibinfo {year} {1983})}\BibitemShut {NoStop}%
\bibitem [{\citenamefont {Vlassis}(2004)}]{Vlassis2004}%
  \BibitemOpen
  \bibfield  {author} {\bibinfo {author} {\bibfnamefont {Jo{\"e}lle}\
  \bibnamefont {Vlassis}},\ }\bibfield  {title} {\enquote {\bibinfo {title}
  {Making sense of the minus sign or becoming flexible in ``negativity''},}\
  }\href {\doibase https://doi.org/10.1016/j.learninstruc.2004.06.012}
  {\bibfield  {journal} {\bibinfo  {journal} {Learning and Instruction}\
  }\textbf {\bibinfo {volume} {14}},\ \bibinfo {pages} {469 -- 484} (\bibinfo
  {year} {2004})}\BibitemShut {NoStop}%
\bibitem [{\citenamefont {Brahmia}\ and\ \citenamefont
  {Boudreaux}(2016)}]{Brahmia2016b}%
  \BibitemOpen
  \bibfield  {author} {\bibinfo {author} {\bibfnamefont {Suzanne~S}\
  \bibnamefont {Brahmia}}\ and\ \bibinfo {author} {\bibfnamefont {Andrew}\
  \bibnamefont {Boudreaux}},\ }\bibfield  {title} {\enquote {\bibinfo {title}
  {Exploring student understanding of negative quantity in introductory physics
  contexts},}\ }in\ \href@noop {} {\emph {\bibinfo {booktitle} {Proceedings of
  the 19th Annual Conference of RUME}}}\ (\bibinfo {year} {2016})\ p.~\bibinfo
  {pages} {79}\BibitemShut {NoStop}%
\bibitem [{\citenamefont {Bajracharya}\ \emph {et~al.}(2012)\citenamefont
  {Bajracharya}, \citenamefont {Wemyss},\ and\ \citenamefont
  {Thompson}}]{Bajracharya2012}%
  \BibitemOpen
  \bibfield  {author} {\bibinfo {author} {\bibfnamefont {Rabindra~R.}\
  \bibnamefont {Bajracharya}}, \bibinfo {author} {\bibfnamefont {Thomas~M.}\
  \bibnamefont {Wemyss}}, \ and\ \bibinfo {author} {\bibfnamefont {John~R.}\
  \bibnamefont {Thompson}},\ }\bibfield  {title} {\enquote {\bibinfo {title}
  {Student interpretation of the signs of definite integrals using graphical
  representations},}\ }\href {\doibase 10.1063/1.3680006} {\bibfield  {journal}
  {\bibinfo  {journal} {AIP Conference Proceedings}\ }\textbf {\bibinfo
  {volume} {1413}},\ \bibinfo {pages} {111--114} (\bibinfo {year}
  {2012})}\BibitemShut {NoStop}%
\bibitem [{\citenamefont {Hayes}\ and\ \citenamefont
  {Wittmann}(2010)}]{Hayes2010}%
  \BibitemOpen
  \bibfield  {author} {\bibinfo {author} {\bibfnamefont {Kate}\ \bibnamefont
  {Hayes}}\ and\ \bibinfo {author} {\bibfnamefont {Michael~C.}\ \bibnamefont
  {Wittmann}},\ }\bibfield  {title} {\enquote {\bibinfo {title} {The role of
  sign in students' modeling of scalar equations},}\ }\href {\doibase
  10.1119/1.3361994} {\bibfield  {journal} {\bibinfo  {journal} {The Physics
  Teacher}\ }\textbf {\bibinfo {volume} {48}},\ \bibinfo {pages} {246--249}
  (\bibinfo {year} {2010})}\BibitemShut {NoStop}%
\bibitem [{\citenamefont {Doran}(1980)}]{Doran1980}%
  \BibitemOpen
  \bibfield  {author} {\bibinfo {author} {\bibfnamefont {Rodney~L.}\
  \bibnamefont {Doran}},\ }\href@noop {} {\emph {\bibinfo {title} {Basic
  Measurement and Evaluation of Science Instruction.}}}\ (\bibinfo  {publisher}
  {National Science Teachers Association},\ \bibinfo {year} {1980})\BibitemShut
  {NoStop}%
\bibitem [{\citenamefont {Adams}\ and\ \citenamefont
  {Wieman}(2010)}]{Adams2010}%
  \BibitemOpen
  \bibfield  {author} {\bibinfo {author} {\bibfnamefont {Wendy~K.}\
  \bibnamefont {Adams}}\ and\ \bibinfo {author} {\bibfnamefont {Carl~E.}\
  \bibnamefont {Wieman}},\ }\bibfield  {title} {\enquote {\bibinfo {title}
  {Development and validation of instruments to measure learning of expert-like
  thinking},}\ }\href@noop {} {\bibfield  {journal} {\bibinfo  {journal}
  {International Journal of Science Education}\ }\textbf {\bibinfo {volume}
  {33}},\ \bibinfo {pages} {1289--1312} (\bibinfo {year} {2010})}\BibitemShut
  {NoStop}%
\bibitem [{\citenamefont {Wiersma}\ and\ \citenamefont
  {Jurs}(1990)}]{wiersma1985educational}%
  \BibitemOpen
  \bibfield  {author} {\bibinfo {author} {\bibfnamefont {William}\ \bibnamefont
  {Wiersma}}\ and\ \bibinfo {author} {\bibfnamefont {Stephen~G}\ \bibnamefont
  {Jurs}},\ }\href@noop {} {\emph {\bibinfo {title} {Educational measurement
  and testing}}},\ \bibinfo {edition} {2nd}\ ed.\ (\bibinfo  {publisher} {Allyn
  \& Bacon},\ \bibinfo {year} {1990})\BibitemShut {NoStop}%
\bibitem [{Note1()}]{Note1}%
  \BibitemOpen
  \bibinfo {note} {Our decision to use the multiple-response, single-correct
  (MRS) format is based on the desire to see whether or not students will
  choose incorrect responses in combination with a single correct response.
  This also allows us to measure students' likelihood to choose various
  combinations of incorrect responses.}\BibitemShut {Stop}%
\end{thebibliography}%

\end{document}